\documentclass[american,prb,twocolumn,superscriptaddress]{revtex4-2}
\usepackage[T1]{fontenc}
\usepackage[latin9]{inputenc}
\usepackage{amsmath}
\usepackage{amssymb}
\usepackage{graphicx}

\makeatletter
\usepackage{times}

\makeatother

\usepackage{babel}
\begin{document}
\title{Dynamic phases induced by two-level system defects on driven qubits}
\author{Yanxiang Wang}
\affiliation{Institute of Applied Physics and Materials Engineering, University
of Macau, Macau, China}
\author{Ziyang You}
\affiliation{Institute of Applied Physics and Materials Engineering, University
of Macau, Macau, China}
\author{Hou Ian }
\affiliation{Institute of Applied Physics and Materials Engineering, University
of Macau, Macau, China}
\homepage{houian@um.edu.mo}

\begin{abstract}
Recent experimental evidences point to two-level defects, located
in the oxides and on the interfaces of the Josephson junctions, as
the major constituents of decoherence in superconducting qubits. How
these defects affect the qubit evolution with the presence of external
driving is less well understood since the semiclassical qubit-field
coupling renders the Jaynes-Cummings model for qubit-defect coupling
undiagonalizable. We analyze the decoherence dynamics in the continuous
coherent state space induced by the driving and solve the master equation
endowed with an extra decay-cladded driving term via a Fokker-Planck
equation. The solutions for diffusion propagators as Gaussian distributions
show four distinct dynamic phases: four types of convergence paths
to limit cycles of varying radius by the distribution mean, which
are determined by the competing external driving and the defect decays.
The qubit trajectory resulted from these solutions is a super-Poissonian
over displac ed Fock states, which reduces to a Gibbs state of effective
temperature decided by the defect at zero driving limit. Further,
the Poincare map shows the dependence of the rate of convergence on
the initial state. In other words, the qubit evolution can serve as
an indicator of the defect coupling strength through the variation
of the driving strength as a parameter.
\end{abstract}
\maketitle

\section{Introduction}

Environmental coupling causes decoherent evolution of qubits. For
superconducting qubits, decoherence is especially acute because of
their inevitable couplings to the surrounding materials in the solid-state
circuits in addition to the vacuum coupling~\citep{Ithier05}. Since
finite times are required for performing quantum logical operations,
improving decoherence times has been a major field of study for more
than a decade~\citep{Krantz19}, with experiments demonstrating $T_{1}$
and $T_{2}$ times on the order of 10$\mu$s~\citep{Peterer15} and
100$\mu$s~\citep{nguyen19}, respectively. Nevertheless, characteristics
of decoherence are often precarious and sometimes highly random in
fabricated superconducting circuits and an analytical understanding
of the sources for decoherence is still required. 

Many studies are directed towards sourcing the material origin of
qubit decoherence~\citep{Muller}, which point to the defects~\citep{Corcoles11,Neill}
located broadly in the interior and on the edges of the sandwiched
oxides of the tunnel barriers as well as chip surfaces~\citep{Lisenfeld15,Meissner}.
Spectroscopic analyses show that the defects, growing out of the amorphous
structure from natural or slow oxidation process, can be modelled
as two-level systems~\citep{Muller}. In particular, these two-level
defects are in general categorized into two types: fluctuators, with
level spacing $\Delta E\lesssim k_{B}T$, and coherent two-level systems
(cTLS), with level spacing $\Delta E\gg k_{B}T$; the former (latter)
are thus strongly (weakly) coupled to the environment at temperature
$T$~\citep{Faoro15}. Collectively, cTLS and fluctuators contribute
a 1/f background noise spectrum~\citep{Faoro06} and their interaction
would decide the power dependence of the quality factor of superconducting
resonators~\citep{Faoro12}. Yet the greater energy gap places cTLS
spectroscopically near the qubit, making the qubit effectively more
susceptible to the motions of the cTLS than those of the fluctuators.
These coherent two-level defects are thus recognized as the dominant
source for relaxing a qubit's coherent dynamics~\citep{Klimov,Abdurakhimov}
and methods to alleviate such relaxations have been proposed~\citep{Matityahu21,You22}.

Recent experimental progress has characterized the qubit relaxations
induced by cTLS~\citep{Schlor,Lu21,Bilmes20,Bejanin21,Bejanin22},
where the theoretical analyses are modeled on the interactions between
Pauli operators (for cTLS) and anharmonic oscillators (for qubits)~\citep{Nakamura99,Koch,Houck07}.
Superconducting qubits including the prevalent transmon variants are
indeed anharmonic multi-level systems whose higher excited states
are often ignored to simplify the discussion of their roles in quantum
computations. However, when relaxations towards defects are taken
into account, the contributions from the higher levels become significant
and must be considered~\citep{Abdurakhimov}. We note that in optical
amorphous materials, the defect couplings are described by non-Hermitian
Hamiltonians~\citep{Ozdemir19}.

Since relaxations occur concurrently with driving and their rates
are highly dependent on the driving strength~\citep{Ian10,Yan,gao20_1,gao20_2,Gu17,Ladd,Smirnov,Geva},
characterizing the decoherence dynamics under driving becomes necessary
and nontrivial because the semiclassical qubit-field coupling removes
the discrete diagonalizability of the qubit-defect coupling. The effective
dressed relaxation rates are first computed in Ref.~\citep{Kustura21}.

Here, we resort to a solution of the evolution on the continuous complex
$(\alpha,\alpha^{\ast})$-plane of coherent states by regarding the
driving as a displacement that translates the creation and annhilation
operator pair of the anharmonic qubit. Since spectral analysis shows
the noise additional to the 1/f background contributed by the fluctuators
stems from a single two-level system, we register a single perturbative
term provided by the cTLS in the master equation. The qubit decoherence
evolution is then fully solved by writing its density matrix in the
$Q$-representation, whence the master equation is replaced by a Fokker-Planck
(FP) equation. The solution $Q(\alpha,\alpha^{\ast},t)$, obtained
by solving two diffusion-type coupled equation of real variables derived
from the FP equation, obeys a Gaussian distribution of moving mean
and expanding variance.

We find that the steady-state of the moving mean follows a limit cycle
determined by both the driving field and the decays induced by the
cTLS. The steady state regarded as a thermal state has the qubit thermal
inversion decided by the cTLS level spacing in addition to the environmental
temperature $T$. Given an initial point state on the quadrature plane,
the radii of the asymptotic limit cycles can be either greater or
less than the distance to origin from that initial point, depending
on both the qubit effective decay mediated by the cTLS and the driving
strength. The transient behaviors towards these limit cycles is even
more sensitive to the driving condition: it can generally be classified
into four dynamic phases of distinct converging behaviors according
to the competing driving and decay. This sensitivity translates into
an indicator of the decay strength induced by the cTLS under the varying
driving strength as a parameter. To understand how these dynamic phases
arise, we begin with the discussion of the tripartite interaction
model below in Sec.~\ref{fig:model}, in which the master equation
is introduced, and follow it by the process of solution in Sec.~\ref{sec:solution}.
The classification of the dynamic phases is presented in Sec.~\ref{sec:classification}.
The transient dynamics is analyzed in Sec.~\ref{sec:transient} before
the conclusions are given in Sec.~\ref{sec:Conclusions}.

\section{Qubit-cTLS coupling under displacements\protect\label{sec:model}}

\begin{figure}
\includegraphics[clip,width=8.5cm]{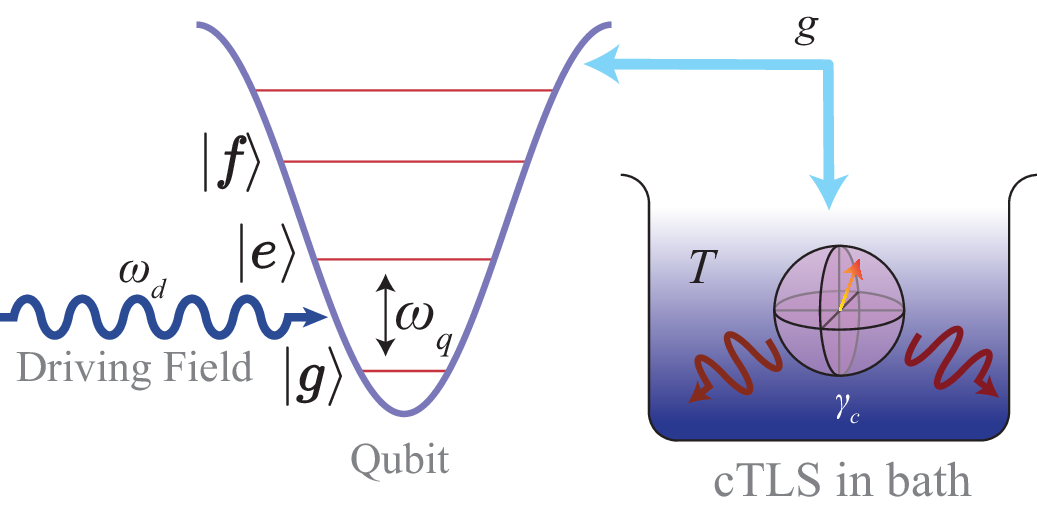}

\caption{The system consists of an anharmonic oscillator of base level spacing
$\omega_{q}$ and anharmonicity $\chi$ that models a superconducting
qubit, a two-level system of spacing $\omega_{c}$ that models a near-resonant
defect in the qubit junction, and a microwave driving of frequency
$\omega_{d}$ to the qubit. The two-level defect is considered interacting
with a thermal bath of temperature $T$ at a decay rate of $\gamma_{c}$
and simultaneously interacting with the qubit at strength $g$. The
latter interaction provides a two-way channel of energy flow during
the qubit thermalization process with the bath.~\protect\label{fig:model}}

\end{figure}

The total Hamiltonian ($\hbar=1$)

\begin{equation}
H=H_{q}+H_{c}+H_{\mathrm{int}}+W\label{eq:raw_Ham}
\end{equation}
has four terms, accounting for the three comprising parts of the system
and the environmental interactions as illustrated in Fig.~\ref{fig:model}.
The first term $H_{q}=\omega_{q}a^{\dagger}a-\chi a^{\dagger2}a^{2}$
denotes the free energy of the qubit with resonance frequency $\omega_{q}$
and anharmonicity $\chi$. The anharmonic oscillator model arises
from the double-well potential that are quantized from the junction
and capacitor energy of the current and voltage in the solid-state
based system~\citep{Clarke,Orlando99}. The second term $H_{c}=\omega_{c}\sigma_{z}/2$
accounts for the free energy of the cTLS. The last two terms denote
the interactions, for the qubit-cTLS coupling $H_{\mathrm{int}}=g\left(a^{\dagger}\sigma_{-}+a\sigma_{+}\right)$
and for the semiclassical field driving $W(t)=\Omega\left(a^{\dagger}e^{-i\omega_{d}t}+\mathrm{h.c.}\right)$,
where $\Omega$ is the driving strength and $\omega_{d}$ the field
frequency.

Considering the typical scenario where $\Omega\gg g$ ($\Omega$ in
the range of MHz while $g$ in the range of hundreds of kHz for typical
transmon qubits), the driving is relatively strong, compared to which
the coupling to cTLS is perturbative. The discrete qubit state space
can then be transformed into the continuous coherent-state space under
the rotating frame of the driving $\omega_{d}$, where $\Omega$ and
the qubit-drive detuning $\Delta_{d}=\omega_{q}^{\prime}-\omega_{d}$
determine the amount of displacement $\alpha_{d}=\Omega/\Delta_{d}$
from the vacuum state, i.e. $H_{q}+W(t)=D(\alpha_{d})H_{q}D^{\dagger}(\alpha_{d})$.
Given weak driving $\Omega$, the magnitude of the displacement is
small, which renders the qubit to fall into the weak-excitation regime~\citep{Wiegand21},
where the oscillator frequency can be linearized as $\omega_{q}^{\prime}=\omega_{q}-\chi(\bar{n}_{q}-1)$.
The number $\bar{n}_{q}$ denotes the average photon number in the
qubit, which remains a constant value since the coupling strengths
$g,\Omega\ll\omega$. By regarding $DH_{q}D^{\dagger}$ and $H_{c}$
as free energies in the coherent-state space, the effective qubit-cTLS
coupling under the interaction picture becomes
\begin{equation}
H_{\mathrm{eff}}=g\left(a^{\dagger}-\alpha_{d}e^{-i\Delta_{d}t}\right)\sigma_{-}e^{i\Delta_{c}t}+\textrm{h.c.}\label{eq:eff_Ham}
\end{equation}
where $\Delta_{c}=\omega_{q}^{\prime}-\omega_{c}$ denotes cTLS detuning
from the linearized qubit. 

Given Eq.~(\ref{eq:eff_Ham}), the master equation of the qubit density
matrix $\rho_{q}$ is derived from the Liouville equation $\dot{\rho}=-i\left[H_{\mathrm{eff}},\rho\right]$
for the coupled system $\rho$. We apply the Born-Oppenheimer-Markov
approximation since the motion of the cTLS is relatively much slower
than the fast motion of the qubit. Experimental studies~\citep{Klimov,Schlor}
have shown that cTLS-induced variations in frequency and dephasing
rate of the qubit occur over the scale of hours to days; the cTLS
evolution is therefore not perturbed by the weak qubit-cTLS coupling
over microsecond time scale considered here. Exactly, the density
matrix is assumed separable, i.e. $\rho(t)=\rho_{q}(t)\otimes\rho_{c}$,
at initial moment before the time integration is carried out to the
first perturbative order and the cTLS subsystem is traced out. As
shown in Appendix A, this leads to
\begin{align}
\frac{\partial}{\partial t}\rho_{q}= & -i\Delta_{d}\left[a^{\dagger}a,\rho_{q}\right]+\eta\left[a^{\dagger}-a,\rho_{q}\right]\nonumber \\
 & +\frac{\gamma}{2}(1-\nu)\left(2a\rho_{q}a{}^{\dagger}-\left\{ a^{\dagger}a,\rho_{q}\right\} \right)\nonumber \\
 & +\frac{\gamma}{2}\nu\left(2a^{\dagger}\rho_{q}a-\left\{ aa^{\dagger},\rho_{q}\right\} \right),\label{eq:master_eqn}
\end{align}
in the displaced $(\alpha,\alpha^{\ast})$-space. Besides the two
Lindbladians contributed by direct coupling to the cTLS, the indirect
coupling skewed by the driving field contributes the extra $\eta$
term. With $\nu=(\exp(\omega_{c}/k_{B}T)+1)^{-1}$ indicating the
thermal inversion of the cTLS, the Lindbladians accounts for the spontaneous
qubit emission (the third term) and absorption (the last term), where
$\gamma=2g^{2}\gamma_{c}/(\gamma_{c}^{2}+\Delta_{c}^{2})$ is the
decay rate according to the typical relaxation rate $\gamma_{c}$
of a cTLS~\citep{Lisenfeld16}. The $\eta$ term reflects the external
driving under the influence of the environment, i.e.
\begin{align}
\eta & =\alpha_{d}g^{2}\left[\frac{(1-\nu)\gamma_{c}}{\gamma_{c}^{2}+(\Delta_{d}+\Delta_{c})^{2}}-\frac{\nu\gamma_{c}}{\gamma_{c}^{2}+(\Delta_{d}-\Delta_{c})^{2}}\right],\label{eq:drive_decay}
\end{align}
which comprises two terms. The first term of Eq.~(\ref{eq:drive_decay})
is contributed by a two-photon process where the driving field photon
and the cTLS radiation photon concurrently mix with the qubit. Whereas,
the second term corresponds to a one-photon process where the driving
field photon bypasses the qubit and mixes with the cTLS. The resonance
at the latter blocks the decay of the qubit that channels into the
cTLS. It therefore adversely affects the decay depicted in Eq.~(\ref{eq:master_eqn})
and has the negative coefficient. On the other hand, both terms share
the common factor $\alpha_{d}$ determined by the driving amplitude.
Given the environmental temperature (on the mK range) of superconducting
qubits, the inversion $v$ adopts a negligible value and the existence
of the driving field in general accelerates qubit relaxations (contribution
of the first term $\gg$ that of the second term).

\section{Solving for quadratures\protect\label{sec:solution}}

To solve the master equation (\ref{eq:master_eqn}) analytically,
we first reduce it to a Fokker-Planck (FP) equation concerning the
density distribution of the qubit on the complex $\alpha$-plane.
Under the $Q$-representation $Q(\alpha,\alpha^{\ast},t)=\left\langle \alpha\right|\rho_{q}(t)\left|\alpha\right\rangle /\pi$,
the FP equation reads
\begin{align}
\frac{\partial}{\partial t}Q= & \left[\left(\bar{\nu}\gamma+i\Delta_{d}\right)\frac{\partial}{\partial\alpha}\alpha-\eta\frac{\partial}{\partial\alpha}+\mathrm{h.c.}\right]Q\nonumber \\
 & +(1-\nu)\gamma\frac{\partial^{2}}{\partial\alpha\partial\alpha^{\ast}}Q,\label{eq:FP_eqn}
\end{align}
where the square bracket indicates an Hermitian differential operator
and $\bar{\nu}=1/2-\nu$ indicates the inversion difference between
$1/2$ at infinite temperate and $\nu$ at a given temperature~\citep{Carmichael13B}.
To arrive at an analytical solution to Eq.~(\ref{eq:FP_eqn}), we
first rewrite the differential equation in the real quadrature plane
under a rotating frame: $x=(\alpha e^{i\Delta_{d}t}+\mathrm{h.c.})/2$
and $y=-i(\alpha e^{i\Delta_{d}t}-\mathrm{h.c.})/2$ (Cf. Appendix
B), giving 
\begin{align}
\frac{\partial}{\partial t}\tilde{Q} & (x,y,t)=\biggl[\bar{\nu}\gamma\left(\frac{\partial}{\partial x}x+\frac{\partial}{\partial y}y\right)\nonumber \\
 & +\frac{(1-\nu)}{4}\gamma\left(\frac{\partial^{2}}{\partial x^{2}}+\frac{\partial^{2}}{\partial y^{2}}\right)\nonumber \\
 & -\eta\left(\cos\Delta_{d}t\frac{\partial}{\partial x}+\sin\Delta_{d}t\frac{\partial}{\partial y}\right)\biggr]\tilde{Q}(x,y,t).
\end{align}
Since the initial distribution $\tilde{Q}(x,y,0)$ is not necessarily
separable along the $x$ and $y$ variables, we consider the associated
equation of motion for the propagator (\emph{viz. }the Green's function)
$\mathcal{G}(x,y,t;x',y',0)$ linking the initial coordinate $(x',y')$
and the current coordinate $(x,y)$ at current time $t$~\citep{Carmichael13B}.
Since it reduces to the separated $\delta(x-x')\delta(y-y')$ at the
time limit $t\to0$, the propagator is separable as $\mathcal{G}_{x}(x,t;x',0)\mathcal{G}_{y}(y,t;y',0)$
in general, breaking up its equation of motion into the pair 
\begin{align}
\frac{\partial\mathcal{G}_{x}}{\partial t} & =\left[\bar{\nu}\gamma+\left(\bar{\nu}\gamma x-\eta\cos\Delta_{d}t\right)\frac{\partial}{\partial x}+\frac{(1-\nu)\gamma}{4}\frac{\partial^{2}}{\partial x^{2}}\right]\mathcal{G}_{x},\label{eq:X_eqn}\\
\frac{\partial\mathcal{G}_{y}}{\partial t} & =\left[\bar{\nu}\gamma+\left(\bar{\nu}\gamma y-\eta\sin\Delta_{d}t\right)\frac{\partial}{\partial y}+\frac{(1-\nu)\gamma}{4}\frac{\partial^{2}}{\partial y^{2}}\right]\mathcal{G}_{y}.\label{eq:Y_eqn}
\end{align}

Each equation above is essentially equivalent to a heat diffusion
equation, where the first-order derivatives and the linear terms on
the right hand side can be eliminated under a new space variable with
the time varible transformed accordingly~\citep{Polyanin15B}. The
diffusion equations are analytically solvable using standard Fourier
transforms (Cf. Appendix C). Eventually, reversing all the transforms
and changes of variables, one obtains the $Q$-distribution at time
$t$ from an initial coherent state distribution under the joint propagator
$\mathcal{G}_{x}\mathcal{G}_{y}$ in the complex plane:
\begin{equation}
Q(\alpha,\alpha^{\ast},t)=\frac{1}{2\pi\sigma^{2}(t)}\exp\left\{ -\frac{\left|\alpha-\mu(t)\right|^{2}}{2\sigma^{2}(t)}\right\} \label{eq:Q-rep}
\end{equation}
which is Gaussian with a moving mean $\mu(t)=\mu_{ss}\exp(i\Delta_{d}t)+(\alpha_{0}-\mu_{ss})\exp(-\bar{\nu}\gamma t)$,
assuming $\alpha_{0}$ as the displacement of the coherent state.
The mean converges to the asymptotic coordinates 
\begin{equation}
\mu_{\mathrm{ss}}=\frac{\eta}{\bar{\nu}\gamma+i\Delta_{d}}-\alpha_{d}\label{eq:ss_mean}
\end{equation}
at steady state, which depends on the external driving as well as
the qubit decay. The variance $\sigma^{2}=\left[1-\exp(-2\bar{\nu}\gamma t)\right]\nu/4\bar{\nu}+1/2$
is expanding and convergent with time.

To verify the validity of the analytical approach, we employ the matrix
exponentiation method to numerically solve the master equation in
a superoperator form~\citep{Weinstein12} and find that the discrepancy
between the analytical $|\mu(t)|$ and the numerical $|\mu(t)|$ is
marginal, a difference on the order of $10^{-4}$ over unital order
of magnitude (Cf. Appendix F for comparison details). 

\section{Classifying evolution\protect\label{sec:classification}}

The evolution depicted by Eq.~(\ref{eq:Q-rep}) dictates that the
qubit, starting from an arbitrary (pure) coherent state $\left|\alpha_{0}\right\rangle $
for which $Q(0)\sim\mathcal{N}(\alpha_{0},1/2)$, follows a typical
trajectory of decay into the quasi-coherent state $\rho_{\mathrm{ss}}$
for which $Q(t\to\infty)\sim\mathcal{N}(\mu_{\mathrm{ss}},\nu/4\bar{\nu}+1/2)$
independent of the initial state $\left|\alpha_{0}\right\rangle $.
The mean $\mu_{\mathrm{ss}}$ depends only on the parameters of the
driving signal and the decay parameters of the cTLS. In fact, if the
driving vanishes, the equilibrium distribution $Q_{\mathrm{ss}}$
reduces to that of an equivalent thermal Gibbs state~\citep{LOUDON},
which reads $\exp\left\{ -|\alpha|^{2}/(\bar{n}+1)\right\} /\pi(\bar{n}+1)$
in the continuous $\alpha$-plane with $\bar{n}=\nu/2\bar{\nu}=(\exp\{\omega_{c}/k_{B}T\}-1)^{-1}$
indicating the equivalent average photon number. In other words, the
qubit levels would be populated not according to $\omega_{q}^{\prime}$,
but rather to the level spacing $\omega_{c}$ of cTLS regarded as
an infinite-level harmonic oscillator. One therefore can follow the
model in Fig.~\ref{fig:model} and regard the qubit-cTLS interaction
as a thermalization channel, for which the qubit is thermalized with
a cTLS-mediated bath at the effective temperature $T_{q}=T\omega_{q}^{\prime}/\omega_{c}$.
Given the ultra-low bath temperature $T$ of the dilution fridge,
the difference in qubit population inversions under $T$ and $T_{q}$
is significant, despite the near-resonant frequencies of $\omega_{q}^{\prime}$
and $\omega_{c}$. This temperature difference is recently measured~\citep{Kulikov20}
and is useful for determining $\omega_{c}$~\citep{Lisenfeld16,Carroll22}.
Moreover, the widened variance of $\sigma_{\mathrm{ss}}^{2}$ by $\nu/4\bar{\nu}$
compared to $\sigma^{2}(0)$ also stems from the thermal energy transferred
from the cTLS to the qubit.

\begin{figure}
\includegraphics[bb=15bp 0bp 245bp 345bp,clip,width=8.6cm]{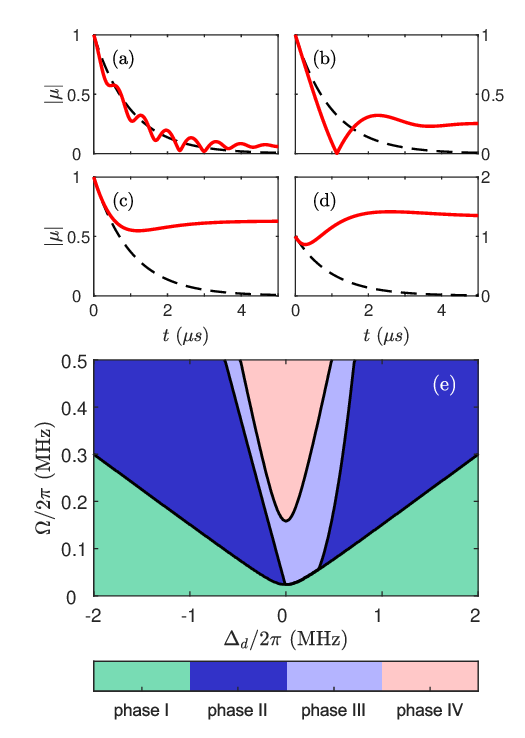}

\caption{Dynamic phases of the decohering evolution of a superconducting qubit
under driving, starting from the same initial point $\alpha_{0}=1$
in the complex $(\alpha,\alpha^{\ast})$-plane. The common parameters
include the qubit frequency $\omega_{q}^{\prime}/2\pi=4.5$~GHz,
the qubit-cTLS detuning $\Delta_{c}/2\pi=110$~kHz, the interaction
strength $g/2\pi=400$~kHz, the cTLS relaxation rate $\gamma_{c}/2\pi=1$~MHz,
and the environmental temperature $T=10$~mK. The dynamic evolution
exhibits distinct behaviors that can be categorized into four phases
under distinct driving detunings $\Delta_{d}/2\pi$ and strengths
$\Omega/2\pi$: (a) $-1.5$~MHz and $100$~kHz for phase I (green);
(b) $-370$~kHz and $100$~kHz for phase II (deep blue); (c) $20$~kHz
and $100$~kHz for phase III (light purple); and (d) $100$~kHz
and $250$~kHz for phase IV (pink). The dashed curves in (a)-(d)
shows the evolution of $\mu(t)$ without driving as a reference. (e)
shows the partitions of these dynamic phases over a range of detunings
and strengths.\protect\label{fig:dynamic_phases}}
\end{figure}

For the scenario of a finite driving strength, the steady-state $Q_{\mathrm{ss}}$
has a mean $\mu_{\mathrm{ss}}$ distant from the origin, whose distribution
is super-Poissonian over the displaced Fock states~\citep{Kral90,Keil11}
$\left|\mu_{\mathrm{ss}},n\right\rangle $, i.e. the corresponding
density matrix is
\begin{equation}
\rho_{q}=\sum_{n=0}^{\infty}\frac{(2\sigma_{\mathrm{ss}}^{2}-1)^{n}}{(2\sigma_{\mathrm{ss}}^{2})^{n+1}}\left|\mu_{\mathrm{ss}},n\right\rangle \left\langle \mu_{\mathrm{ss}},n\right|,\label{eq:super-Poisson}
\end{equation}
which would fall back to the Gibbs state described above when the
driving vanishes. When the driving is present, the transient evolutions
of $Q(\alpha,\alpha^{\ast},t)$ under different driving strengths
and frequencies towards $Q_{\mathrm{ss}}$ are not unified but classifiable
into four dynamic categories or phases (Cf. Appendix D): (I) oscillating,
(II) collapse and revive, (III) decay and revive, and (IV) amplifying.
These phases demonstrate the competition of the external driving as
a restoring energy flow against the qubit decay into the cTLS during
the decohering evolution. The typical plots for these four phases
of evolutions are shown in Fig.~\ref{fig:dynamic_phases}, along
with the phase partitions over the qubit-driving detuning $\Delta_{d}$
and the driving strength $\Omega$. At near resonance, the driving
becomes more effective in pushing the steady-state $\mu_{\mathrm{ss}}$
away from zero. According to the evolution of the $Q$-distribution,
the energy $\left\langle H_{q}(t)\right\rangle =\hbar\omega_{q}^{\prime}(|\mu(t)|^{2}+2\sigma^{2}(t)-1)$
of the qubit varies with time in a way that if the driving is weak
or far-resonant such that $|\mu_{\mathrm{ss}}|<|\alpha_{0}|$, the
energy supplied by the driving to the qubit is not sufficient to cover
the energy loss towards the cTLS (phase I to III). At near resonance
with sufficiently large driving, however, the qubit can accumulate
energy in addition to dissipating into the cTLS, exhibiting the scenario
of phase IV. Further analysis of the phase partitions are given in
Appendix D.

\section{Transient dynamics\protect\label{sec:transient}}

\begin{figure}
\includegraphics[bb=0bp 0bp 260bp 436bp,clip,width=8cm]{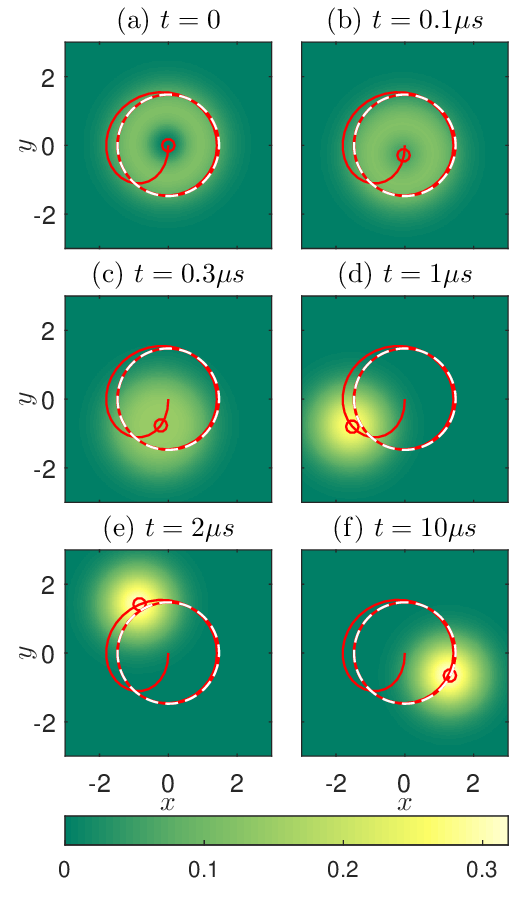}

\caption{The evolution of the qubit $Q$-distribution plotted as contours on
\emph{XY}-quadrature plane, assuming a fully inverted excited state$\left|e\right\rangle $
as the initial state. The captures of six time moments from $t=0$
to $t=10\mu$s are shown in the subplots (a)-(f), where the red curves
show the trajectory of the mean $\mu(t)$ and the overlay white dashed
curve indicates the limit cycle. The driving field parameters are
set to $\Omega/2\pi=500$~kHz and $\Delta_{d}=-300$~kHz, where
the other parameters remain the same as those in Fig.~2.\protect\label{fig:Q-dist_evolution}}
\end{figure}

To appreciate the transient evolution of the qubit under the influence
of the decohering cTLS, we consider the fully inverted $\left|e\right\rangle \left\langle e\right|$
as the qubit initial state. Its $Q$-representation $\left|\left\langle e|\alpha\right\rangle \right|^{2}/\pi$
is the continuous complex function $e^{-|\alpha|^{2}}|\alpha|^{2}/\pi$
on the $(\alpha,\alpha^{*})$ plane, which exhibits a circular-symmetric
ring shape in its contour plot over the observable \emph{$XY$-}quadrature
plane shown in Fig.~\ref{fig:Q-dist_evolution}(a). The quadrature
operators $x$ and $y$ being linear combinations of $a$ and $a^{\dagger}$,
the distribution of $Q(x,y)$ shown in the plot is identical to $Q(\alpha,\alpha^{\ast})$
up to a proportional constant stemmed from the Jacobian matrix. By
shrinking the initial coherent-state distribution of Eq.~\ref{eq:Q-rep}
to a point source, i.e. letting $\sigma_{\mathcal{G}}^{2}(t)=\left[1-\exp(-2\bar{\nu}\gamma t)\right](\nu/4\bar{\nu}+1/2)$,
the solution expressed by Eq.~\ref{eq:Q-rep} represents directly
a propagator $\mathcal{G}(\alpha,t;\alpha_{0},0)$ of the diffusion
process from that point source. Then, the time function written as
the integral $\int d^{2}\alpha_{0}\,|\alpha_{0}|^{2}\exp\{-|\alpha_{0}|^{2}\}\mathcal{G}(\alpha,t;\alpha_{0},0)/\pi$
describes the entire decoherence process (Cf. Appendix E), the illustration
of which at moments after $t=0$ are given in Figs.~\ref{fig:Q-dist_evolution}(b)-(f).
The center of the distribution, marked by a red circle, shows a converging
trajectory (red solid curve) towards a limit cycle of radius $|\mu_{\mathrm{ss}}|$,
marked as a white dashed circle. This shows that after the dynamic
balance between the driving and the energy decay to the cTLS is reached,
the qubit coordinates follow the circular path at the frequency determined
by the detuning $\Delta_{d}$. During the convergence, the ring-shaped
distribution of variance $3\sqrt{\pi}/4$ contracts into a disk shape
of variance $\nu/4\bar{\nu}+1/2$. In other words, the decoherence
process is reflected as a displacement process on the quadrature plane
from the origin to $|\mu_{\mathrm{ss}}|$, where the excited state
distribution decays into the super-Poissonian displaced Fock state
of Eq.~(\ref{eq:super-Poisson}). 

\begin{figure}
\includegraphics[bb=0bp 0bp 305bp 290bp,clip,width=8.3cm]{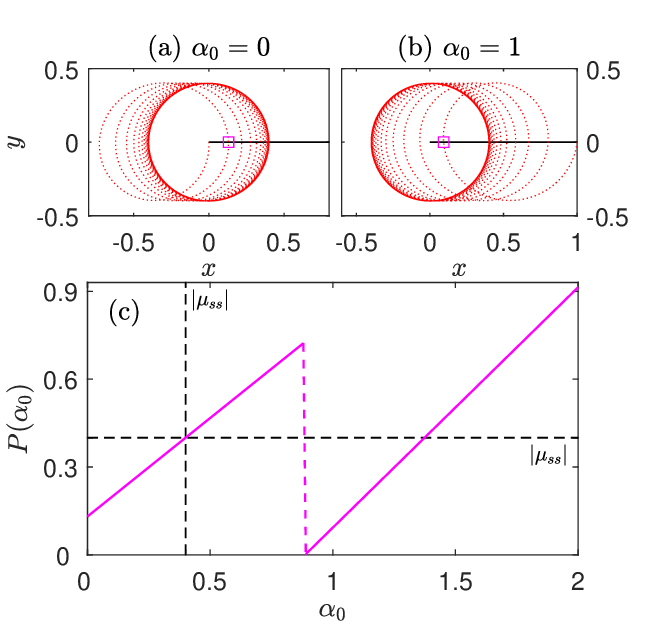}

\caption{Trajectories of $\mu(t)$ drawn as red dotted curves on the $XY$-quadrature
plane, starting from two initial points: (a) $\alpha_{0}=0$ and (b)
$\alpha_{0}=1$, but under the same coupling strength $\Omega/2\pi=1$MHz
and detuning $\Delta=-2.5$~MHz. The solid circles indicate the same
asymptotic limit cycle of radius $|\mu_{\mathrm{ss}}|$ from the distinct
initial points. The magenta squares indicate the first recurrence
on the positive $x$-axis. (c) Poincare map showing the coordinate
of the first recurrence as a function of initial $x$-coordinate $\alpha_{0}$.
The black dashed cross shows the fixed point $\alpha_{0}=0.40$ of
the recurrence function $P(\alpha_{0})=\alpha_{0}=|\mu_{\mathrm{ss}}|$.
The function has two rate zones, one before and one after $\alpha_{0}=0.89$.\protect\label{fig:Poincare_map}}
\end{figure}

Not only does the driving strength $\eta$ under relaxation and the
detuning $\Delta_{d}$ determine the radius of the limit cycle orbit,
they also decide the rate at which the qubit state gravitates towards
it. Shown in Figs.~\ref{fig:Poincare_map}(a)-(b), the trajectories
of $\mu(t)$ converge to the same limit cycle under identical driving
conditions, but approach it under different rates when commencing
from initial points of differing distance to the limit cycle. The
illustrated case of $\alpha_{0}=0$ has a faster approach than that
of $\alpha_{0}=1$. The generalized description of the distinct approaching
rates is given conventionally by the Poincare map, shown in Fig.~\ref{fig:Poincare_map}(c),
where the first recurrence on the positive $x$-axis is recorded as
a function $P(\alpha_{0})$ of the starting coordinate $\alpha_{0}$
on the same axis. The faster rate possessed by the starting coordinate
$\alpha_{0}=0$ is associated with the situation that $P(0)$ is closer
to the fixed point $P(\alpha_{0})=\alpha_{0}$, i.e. $x$ coordinate
of the limit cycle, than $P(1)$. Moreover, the driving conditions
determine a partition point, shown as $\alpha_{0}=0.89$ in the case
illustrated in Fig.~\ref{fig:Poincare_map}(c), separating a zone
of first recurrence occurring at a full circle from another of first
occurrence at a half circle. The former has the $P(\alpha_{0})$ function
increase at a slope 0.67 while the latter 0.82.

\section{Conclusions\protect\label{sec:Conclusions}}

We obtain an analytic description of the evolution of a superconducting
qubit under the simultaneous influence of a coherent two-level system
(cTLS) reflecting the role of a junction defect in a thermal bath
and of an external microwave driving field. By solving a master equation
in the continuous $(\alpha,\alpha^{\ast})$-plane of coherent state,
we has described the thermalization process undergone by the qubit
according to an effective temperature determined by the level spacing
of the cTLS relative to that of the qubit. The strength of the driving
competes with the couplings of the cTLS to the qubit and the environment
to determine the effective qubit relaxation, which are classifiable
into four phases of distinct dynamic behaviors. The competition has
the qubit converge to a limit cycle of finite radius on the quadrature
plane, where the rate of convergence is shown through a Poincare map.
We also note that a more general scenario would arise when the relaxation
is induced by multiple cTLS. The dynamics of the qubit becomes more
intricate and it would be a subject of investigation in future research
endeavors.

\section*{Acknowledgments}

H.I. thanks the support by FDCT of Macau under Grants 0130/2019/A3,
0015/2021/AGJ and 006/2022/ALC.

\section*{APPENDIX A: master equation of qubit-cTLS evolution}

We derive here the master equation~(3) from the Liouville equation
in the Schrödinger picture with the effective Hamiltonian given in
Eq.~(2). It is first derived in the interaction picture before switching
back to the present form under the Schroedinger picture. To be exact,
we consider the density matrix $\rho^{I}(t)=e^{iH_{0}^{D}t}\rho(t)e^{-iH_{0}^{D}t}$
in the interaction picture with the free-energy Hamiltonian $H_{0}^{D}=\frac{1}{2}\omega_{c}\sigma_{z}+\Delta_{d}a^{\dagger}a$
given under the rotating frame $U=\exp(-ia^{\dagger}a\omega_{d}t)$
and the transformation of displacement $D(\alpha_{d})=\exp(\alpha_{d}a^{\dagger}-\alpha_{d}^{*}a)$. 

The evolution of the density matrix follows the Liouville equation
$\partial\rho^{I}(t)/\partial t=-i\left[H_{\mathrm{eff}}(t),\rho^{I}(t)\right]$
in the interaction picture. Because the coupling in the qubit-cTLS
system is weak, the RHS of the equation is expanded up to the second-order
terms under the Born approximation. Besides, we assume the initial
state between the system and the environment be uncorrelated, i.e.
$\rho(t)=\rho_{q}(t)\otimes\rho_{c}(0)$, so that the first-order
terms becomes zero. Further, under the Markov approximation, the short
memory assumption ensures that it is valid to extend the upper limit
of the integral in the iterative term to infinity. Therefore, we arrive
at the dynamics equation of the qubit-reduced density operator 

\begin{align}
\frac{\partial}{\partial t}\rho_{q}^{I}(t) & =\nonumber \\
-\int_{0}^{\infty} & dt'\textrm{tr}_{c}\left[H_{\mathrm{eff}}(t),\left[H_{\mathrm{eff}}(t-t'),\rho_{q}^{I}(t)\otimes\rho_{c}^{I}\right]\right].\tag{A1}\label{eq:B.M.A.}
\end{align}
This equation can be further simplified by expanding the commutator
and tracing out the degree of freedom in cTLS, viz.

\begin{align}
\frac{\partial}{\partial t}\rho_{q}^{I} & =-\left\{ \gamma_{+}\left[a^{\dagger},a\rho_{q}^{I}\right]+\eta_{+}\left[\rho_{q}^{I},a^{\dagger}\right]\right.\nonumber \\
 & \left.+\gamma_{-}\left[a,a^{\dagger}\rho_{q}^{I}\right]+\eta_{-}\left[\rho_{q}^{I},a\right]\right\} +\mathrm{h.c.~},\tag{A2}\label{eq:master_ori}
\end{align}
where the rate coefficients are given by
\begin{align}
\gamma_{\pm}= & g^{2}\int_{0}^{\infty}dt'\left\langle \sigma_{\mp}\sigma_{\pm}\right\rangle e^{\pm i\Delta_{c}t'},\tag{A3}\\
\eta_{\pm}= & \alpha_{d}g^{2}e^{\pm i\Delta_{d}t}\int_{0}^{\infty}dt'\left\langle \sigma_{\mp}\sigma_{\pm}\right\rangle e^{i(\Delta_{d}\pm\Delta_{c})t'}.\tag{A4}
\end{align}
 According to\textbf{ }the Fermi's golden rule, the coefficients $\gamma_{\pm}$
are characterized from the cTLS power spectral density $S(\omega)$
at $\omega=\omega_{q}^{\prime}$ by~\citep{Shnirman05,YouXY21}
\begin{equation}
\gamma_{\pm}=g^{2}\frac{\gamma_{c}\nu_{\pm}}{\gamma_{c}^{2}+\Delta_{c}^{2}},\tag{A5}
\end{equation}
where $\nu_{+}=1-\nu$ and $\nu_{-}=\nu$ with $\nu=1/\left[1+\exp\left(\omega_{c}/T\right)\right]$
being the thermal inversion at bath temperature $T$, and $\gamma_{c}$
is relaxation rate of the cTLS. $\Delta_{c}(\Delta_{d})$ is the qubit
detuning from the cTLS (the driving). Similarly, the coefficients
$\eta_{\pm}$ are given by
\begin{equation}
\eta_{\pm}=\alpha_{d}g^{2}e^{\pm i\Delta_{d}t}\frac{\gamma_{c}v_{\pm}}{\gamma_{c}^{2}+\left(\Delta_{d}\pm\Delta_{c}\right)^{2}}.\tag{A6}
\end{equation}

Substituting these rate coefficients into Eq.~(\ref{eq:master_ori}),
we have
\begin{align}
\frac{\partial}{\partial t}\rho_{q}^{I}= & \frac{\gamma}{2}\left(1-\nu\right)\left(\left[a,\rho_{q}^{I}a{}^{\dagger}\right]+\left[a\rho_{q}^{I},a{}^{\dagger}\right]\right)\nonumber \\
 & +\frac{\gamma}{2}\nu\left(\left[a^{\dagger},\rho_{q}^{I}a\right]+\left[a^{\dagger}\rho_{q}^{I},a\right]\right)\nonumber \\
 & +\eta\left(e^{i\Delta_{d}t}\left[a^{\dagger},\rho_{q}^{I}\right]-e^{-i\Delta_{d}t}\left[a,\rho_{q}^{I}\right]\right),\tag{A7}\label{eq:master_INT}
\end{align}
where $\gamma=\gamma_{+}/\nu_{+}+\gamma_{-}/\nu_{-}$ and $\eta=\eta_{+}e^{-i\Delta_{d}t}-\eta_{-}e^{i\Delta_{d}t}$.
The first parenthesis of Eq.~(\ref{eq:master_INT}) describes the
process where an energy quantum is emitted from the qubit into the
cTLS, while the second describes the reverse process. The parenthesis
on the second line reveals the effect of the driving. Transforming
the master equation back to the Schrödinger picture, one can get the
master equation in the rotating displaced frame:
\begin{align}
\frac{\partial}{\partial t}\rho_{q}= & -i\Delta_{d}\left[a^{\dagger}a,\rho_{q}\right]+\eta\left[a^{\dagger}-a,\rho_{q}\right]\nonumber \\
 & +\frac{\gamma}{2}\left(1-\nu\right)\left(\left[a,\rho_{q}a{}^{\dagger}\right]+\left[a\rho_{q},a{}^{\dagger}\right]\right)\nonumber \\
 & +\frac{\gamma}{2}\nu\left(\left[a^{\dagger},\rho_{q}a\right]+\left[a^{\dagger}\rho_{q},a\right]\right).\tag{A8}\label{eq:master_SCH}
\end{align}

If Born-Oppenheimer-Markov approximation is not considered, the evolution
of the density matrix $\rho^{I}(t)$ follows the Liouville equation
$\partial\rho^{I}(t)/\partial t=-i\left[H_{\mathrm{eff}}(t),\rho^{I}(t)\right]$
where $\rho^{I}(0)=\rho_{q}^{I}(0)\otimes\rho_{c}^{I}(0)$ is separable
at the initial moment. Substituting Eq.~(\ref{eq:eff_Ham}) and applying
$\sigma_{-}e^{i\Delta_{c}t}\to\sigma_{-}$ to simplify the expression,
the equation reads 
\begin{equation}
\frac{\partial\rho^{I}}{\partial t}=-ig\left[a^{\dagger}\sigma_{-},\rho^{I}\right]+ig\alpha_{d}\left[\sigma_{-}e^{-i\Delta_{d}t},\rho^{I}\right]+\textrm{h.c.}\tag{A9}
\end{equation}
Compared to Eq.~(\ref{eq:master_INT}) with the approximation, the
first rerm on RHS corresponds to the Lindbladians with coefficient
$\gamma$ when the partial trace over the cTLS is taken. The second
term corresponds to the external driving with coefficient $\eta$.
Due to the cTLS state remaining approximately unchanged during the
evolution of the qubit~\citep{Schlor}, we can approximate $\rho_{c}^{I}(t)=\rho_{c}^{I}(0)$,
justifying the application of the approximation.

\section*{APPENDIX B: Fokker-Planck equation}

Equations~(7) and (8) are the equations of motion for the real and
imaginary quadratures of the $Q$ representation of the density matrix
$\rho_{q}$ that follows from Eq.~(3). This $Q$ representation of
the qubit is the quasi-probability distribution $Q(\alpha,\alpha^{\ast},t)=\left\langle \alpha\right|\rho_{q}\left|\alpha\right\rangle /\pi$
in the complex $(\alpha,\alpha^{\ast})$-plane~\citep{Carmichael13B}.
Taking the inner product of $\rho_{q}$ according to the definition
on both sides of Eq.~(3) and using the formulae $a\left|\alpha\right\rangle =\alpha\left|\alpha\right\rangle $,
$a^{\dagger}\left|\alpha\right\rangle =(\alpha^{\ast}+\partial/\partial\alpha)\left|\alpha\right\rangle $,
and their Hermitean conjugates, the master equation is rendered into
the Fokker-Planck (FP) equation
\begin{align}
\frac{\partial}{\partial t}Q= & \left[\left(\frac{\gamma}{2}\left(1-2\nu\right)+i\Delta_{d}\right)\frac{\partial}{\partial\alpha}\alpha-\eta\frac{\partial}{\partial\alpha}+\textrm{h.c.}\right]Q\nonumber \\
 & +\gamma\left(1-\nu\right)\frac{\partial^{2}}{\partial\alpha\partial\alpha^{\ast}}Q,\tag{B1}\label{eq:Q_FP}
\end{align}
about the time evolution of the probability density function $Q$.
This type of equations is usually solved by separating it into a system
of two linear equations for the real and the imaginary quadratures.
To do so, we transform the complex variables as $\alpha=\tilde{\alpha}e^{-i\Delta_{d}t}$
and $\alpha^{\ast}=\tilde{\alpha}^{\ast}e^{i\Delta t}$ to eliminate
the imaginary coefficients in the differential operator on the first
line. The FP equation can now be rewritten as
\begin{align}
\frac{\partial}{\partial t}\tilde{Q}= & \frac{\gamma}{2}\left(1-2\nu\right)\left(\frac{\partial}{\partial\tilde{\alpha}}\tilde{\alpha}+\frac{\partial}{\partial\tilde{\alpha}^{\ast}}\tilde{\alpha}^{\ast}\right)\tilde{Q}\nonumber \\
 & +\gamma\left(1-\nu\right)\frac{\partial^{2}}{\partial\tilde{\alpha}\partial\tilde{\alpha}^{\ast}}\tilde{Q}\nonumber \\
 & -\left(\eta e^{i\Delta_{d}t}\frac{\partial}{\partial\tilde{\alpha}}+\mathrm{h.c.}\right)\tilde{Q},\tag{B2}\label{eq:FP_separable}
\end{align}
about $\tilde{Q}(\tilde{\alpha},\tilde{\alpha}^{\ast},t)$, where
the complex coefficients only appear as the oscillating terms on the
second line. Recognizing the real quadratures $x=\Re(\tilde{\alpha})$
and $y=\Im(\tilde{\alpha})$ and consequently rewriting the differential
operators such as 
\begin{equation}
\frac{\partial}{\partial\tilde{\alpha}}=\frac{1}{2}\left(\frac{\partial}{\partial x}-i\frac{\partial}{\partial y}\right),\tag{B3}
\end{equation}
we obtain Eq.~(6) about $\tilde{Q}(x,y,t)$, where all imaginary
coefficients have been eliminated. In general, the $\tilde{Q}$ function
in Eq.~(6) is not separable into a product of functions of $x$ and
$y$ alone, e.g. for the excited state, $\tilde{Q}=(\left|x\right|^{2}+\left|y\right|^{2})\exp(-\left|x\right|^{2}-\left|y\right|^{2})/\pi$
is inseparable. The initial value problem constituted by Eq.~(6)
and an initial condiction $\tilde{Q}(x,y,0)$ can not solved by separation
of variables. Nevertheless, the equation of the propagator associated
with Eq.~(6) can be solved by separation of variables, regarding
the initial condition as a point source delta function. With the solution
of the propagator, i.e. a Green's function $\mathcal{G}(x',y',t;x',y',0)$,
the solution of $\tilde{Q}$ can then be simply obtained by integrating
the product integrand $\mathcal{G}(x',y',t;x',y',0)\tilde{Q}(x',y',0)$
over $x'$ and $y'$.

\section*{APPENDIX C: Solving the Fokker-Planck equation}

The equation associating with Eq.~(6) for the propagator during the
time duration of $0$ to $t$ is~\citep{Carmichael13B}
\begin{align}
\frac{\partial}{\partial t}\mathcal{G} & (x,y,t;x',y',0)\nonumber \\
= & \biggl[\bar{\nu}\gamma\left(\frac{\partial}{\partial x}x+\frac{\partial}{\partial y}y\right)+\frac{(1-\nu)}{4}\gamma\left(\frac{\partial^{2}}{\partial x^{2}}+\frac{\partial^{2}}{\partial y^{2}}\right)\nonumber \\
 & -\eta\left(\cos\Delta_{d}t\frac{\partial}{\partial x}+\sin\Delta_{d}t\frac{\partial}{\partial y}\right)\biggr]\mathcal{G}(x,y,t;x',y',0),\tag{C1}\label{eq:Green_eqn}
\end{align}
where the initial condition is the separated $\lim_{t\to0}\mathcal{G}=\delta(x-x')\delta(y-y')$
corresponding to the point source $\delta(\alpha-\alpha')$ in the
complex plane. When assuming $\mathcal{G}(x,y,t;x',y',0)=\mathcal{G}_{x}(x,t;x',0)\mathcal{G}_{y}(y,t;y',0)$,
one separates Eq.~(\ref{eq:Green_eqn}) into
\begin{align}
\frac{\partial\mathcal{G}_{x}}{\partial t} & =\left[\bar{\nu}\gamma+\left(\bar{\nu}\gamma x-\eta\cos\Delta_{d}t\right)\frac{\partial}{\partial x}+\frac{(1-\nu)\gamma}{4}\frac{\partial^{2}}{\partial x^{2}}\right]\mathcal{G}_{x},\tag{C2}\label{eq:Green_eqn_x}\\
\frac{\partial\mathcal{G}_{y}}{\partial t} & =\left[\bar{\nu}\gamma+\left(\bar{\nu}\gamma y-\eta\sin\Delta_{d}t\right)\frac{\partial}{\partial y}+\frac{(1-\nu)\gamma}{4}\frac{\partial^{2}}{\partial y^{2}}\right]\mathcal{G}_{y}.\tag{C3}\label{eq:Green_eqn_y}
\end{align}
With the help of a transformation of variables~\citep{Polyanin15B},
\begin{align}
\tau & =\frac{1}{2\gamma\bar{\nu}}e^{2\gamma\bar{\nu}t},\tag{C4}\label{eq:t2tau}\\
\xi & =\left(x-\eta\frac{\gamma\bar{\nu}\cos\Delta_{d}t+\Delta_{d}\sin\Delta_{d}t}{\gamma^{2}\bar{\nu}^{2}+\Delta_{d}^{2}}\right)e^{\gamma\bar{\nu}t}\tag{C5}\label{eq:x2ksi}
\end{align}
with $\tau'=\tau|_{t=0}$ and $\xi'=\xi|_{x=x',t=0}$, the Green's
function in the $x$-quadrature can be written as
\begin{equation}
\mathcal{G}_{x}(x,t;x^{\prime},0)=e^{\gamma\bar{\nu}t}\mathcal{G}_{\xi}(\xi,\tau;\xi^{\prime},\tau^{\prime}),\tag{C6}\label{eq:X_to_Ksi}
\end{equation}
for which Eq.~(\ref{eq:Green_eqn_x}) becomes the standard diffusion
equation
\begin{equation}
\frac{\partial\mathcal{G}_{\xi}(\xi,\tau;\xi^{\prime},\tau^{\prime})}{\partial\tau}=\frac{\gamma}{4}\left(\bar{\nu}+1/2\right)\frac{\partial^{2}\mathcal{G}_{\xi}(\xi,\tau;\xi^{\prime},\tau^{\prime})}{\partial\xi^{2}}.\tag{C7}\label{eq:Ksi_Eq.}
\end{equation}

The equation admits the typical Gaussian function solution
\begin{align}
\mathcal{G}_{\xi}( & \xi,\tau;\xi^{\prime},\tau^{\prime})\nonumber \\
= & \frac{1}{\sqrt{\gamma(\bar{\nu}+1/2)(\tau-\tau^{\prime})\pi}}\exp\left\{ -\frac{(\xi-\xi^{\prime})^{2}}{\gamma(\bar{\nu}+1/2)(\tau-\tau^{\prime})}\right\} .\tag{C8}
\end{align}
After substituting Eqs.~(\ref{eq:t2tau}) and (\ref{eq:x2ksi}) into
it, one obtains
\begin{align}
\mathcal{G}_{x}(x,t;x^{\prime},0)= & \frac{1}{\sqrt{2\sigma_{\mathcal{G}}^{2}\pi}}\exp\biggl\{-\Bigl[x-\eta^{\prime}\cos(\Delta_{d}t-\phi)\nonumber \\
 & +(\eta^{\prime}\cos\phi-x^{\prime})e^{-\gamma\bar{\nu}t}\Bigl]^{2}/2\sigma_{\mathcal{G}}^{2}\biggl\},\tag{C9}\label{eq:Green_x}
\end{align}
where $\eta^{\prime}=\eta/\sqrt{\gamma^{2}\bar{\nu}^{2}+\Delta_{d}^{2}}$,
the angle $\phi=\arcsin\left(\Delta_{d}/\sqrt{\gamma^{2}\bar{\nu}^{2}+\Delta_{d}^{2}}\right)$,
and the variance $\sigma_{\mathcal{G}}^{2}=(1/2+\nu/4\bar{\nu})\left(1-\exp(-2\bar{\nu}\gamma t)\right)$.
Similarly, for Eq.~(\ref{eq:Green_eqn_y}), one obtains 
\begin{align}
\mathcal{G}_{y}(y,t;y^{\prime},0)= & \frac{1}{\sqrt{2\sigma_{\mathcal{G}}^{2}\pi}}\exp\biggl\{-\Bigl[y-\eta^{\prime}\sin(\Delta_{d}t-\phi)\nonumber \\
 & -(\eta^{\prime}\sin\phi+y^{\prime})e^{-\gamma\bar{\nu}t}\Bigl]^{2}/2\sigma_{\mathcal{G}}^{2}\biggl\}.\tag{C10}\label{eq:Green_y}
\end{align}
Multiplying Eq.~(\ref{eq:Green_x}) by Eq.~(\ref{eq:Green_y}) results
in the Green's function solution to Eq.~(\ref{eq:Green_eqn}):
\begin{align}
\mathcal{G}(x,y,t;x',y',0) & =\mathcal{G}(\alpha,t;\alpha',0)\nonumber \\
 & =\frac{1}{2\sigma_{\mathcal{G}}^{2}\pi}\exp\left\{ -\frac{\left|\alpha-\mu_{\mathcal{G}}\right|^{2}}{2\sigma_{\mathcal{G}}^{2}}\right\} ,\tag{C11}\label{eq:Green_soln}
\end{align}
which retains the form of Gaussian distribution with moving mean $\mu_{\mathcal{G}}=\mu_{\mathrm{ss}}\exp\left\{ i\Delta_{d}t\right\} +(\alpha^{\prime}-\mu_{\textrm{ss}})\exp\left\{ -\gamma\bar{\nu}t\right\} $
in the complex plane, where $\mu_{\mathrm{ss}}=\eta/(\bar{\nu}\gamma+i\Delta_{d})-\alpha_{d}$
is the asymptotic mean at steady state. Given Eq.~(\ref{eq:Green_soln}),
the distribution $Q$ at any arbitrary time $t$ becomes 
\begin{equation}
Q(\alpha,\alpha^{\ast},t)=\int d^{2}\alpha^{\prime}\mathcal{G}(\alpha,t;\alpha^{\prime},0)Q(\alpha^{\prime},\alpha^{\prime*},0),\tag{C12}\label{eq:Q-Green}
\end{equation}
as explained in the last section.

In particular, if the initial state is a coherent state $\left|\alpha_{0}\right\rangle $,
whose $Q$-distribution is $\exp\left\{ -\left|\alpha-\alpha_{0}\right|^{2}\right\} /\pi$,
the Gaussian Eq.~(9) would arise from Eq.~(\ref{eq:Q-Green}). Compared
to the variance $\sigma_{\mathcal{G}}$ of the propagator, the variance
from this coherent state follows as $\sigma^{2}(t)=\exp\{-2\gamma\bar{\nu}t\}/2+\sigma_{\mathcal{G}}^{2}$
while the mean $\mu(t)=\mu_{\textrm{ss}}\exp\left\{ i\Delta_{d}t\right\} +(\alpha_{0}-\mu_{\textrm{ss}})\exp\left\{ -\bar{\nu}\gamma t\right\} $
is the same of that of the propagator except that the arbitrary $\alpha^{\prime}$
becomes the coherent state displacement $\alpha_{0}$.

\section*{APPENDIX D: Determining dynamical phases of evolution}

We analyze the dynamical features reflected in the evolutions of $Q(\alpha,\alpha^{\ast},t)$,
which are used to obtain the phase diagram Fig.~\ref{fig:dynamic_phases}(e).
This $Q$ function described by Eq.~(9) is a Gaussian distribution
of time-dependent mean $\mu(t)$ and variance $\sigma(t)$. Its general
features such as its envelope rate of convergence and asymptotic value
are characterized by the moving mean. Starting from the same initial
point\textbf{ $\alpha_{0}=1$}, these features become distinct according
to two key parameters: (i) the qubit-driving detuning $\Delta_{d}$
and (ii) the driving strength $\Omega$. We categorize the distinction
as four dynamic phases shown in Fig.~\ref{fig:dynamic_phases}.

Phase I occurs when the driving strength is so weak that the energy
pumped into the qubit by the driving is not sufficient to cover the
energy lost to the cTLS and its environment. Given the finite detuning,
the qubit shows an oscillating $|\mu(t)|$ from $t=0$ to the end
of simulation about the reference curve (plotted with no driving).

When the driving increases to a sufficient strength, the rate of energy
pumped by the driving overcomes the energy loss rate so that asymptotically
$|\mu_{\mathrm{ss}}|$ converges to a finite value at steady state,
giving rise to either the dynamic Phase II or the dynamic Phase III.
Whether the qubit follows the collapse and revival path (II) or the
decay path (III) depends on the detuning $\Delta_{d}$. At close resonance
with small $\Delta_{d}$, the driving is efficient enough to overcome
the energy loss early on so that the qubit trajectory converges directly
into the limit cycle without letting the mean $\mu(t)$ ever reaching
the center of the quadrature plane. In contrast, at large detuning
of $\Delta_{d}$ and thus less effective driving, the qubit is halfway
between the oscillating Phase I and the decay and reviving Phase III,
permitting the $|\mu(t)|$ to hit the zero level when the trajectory
oscillates below the reference damping curve during the first oscillating
period.

Figure \ref{fig:dynamic_phases} shows the cases of trajectories at
negative detuning. When the detuning is positive, the behaviors are
similar, except that the switched sign inverts the dynamic phase of
the trajectories. Consequently, the trajectories undergo an extra
$\pi$-phase worth of oscillation before exhibiting the typical scenarios
of Phases II and III, as illustrated in Fig.~\ref{fig:trajectory}
below. For instance, shown in Fig.~\ref{fig:trajectory}(a), the
collapse appears later in time and, shown Fig.~\ref{fig:trajectory}(b),
there is a bump before the asymptotic value is reached, compared to
the negative detuning cases.

\begin{figure}
\includegraphics[bb=5bp 5bp 310bp 360bp,clip,width=8.5cm]{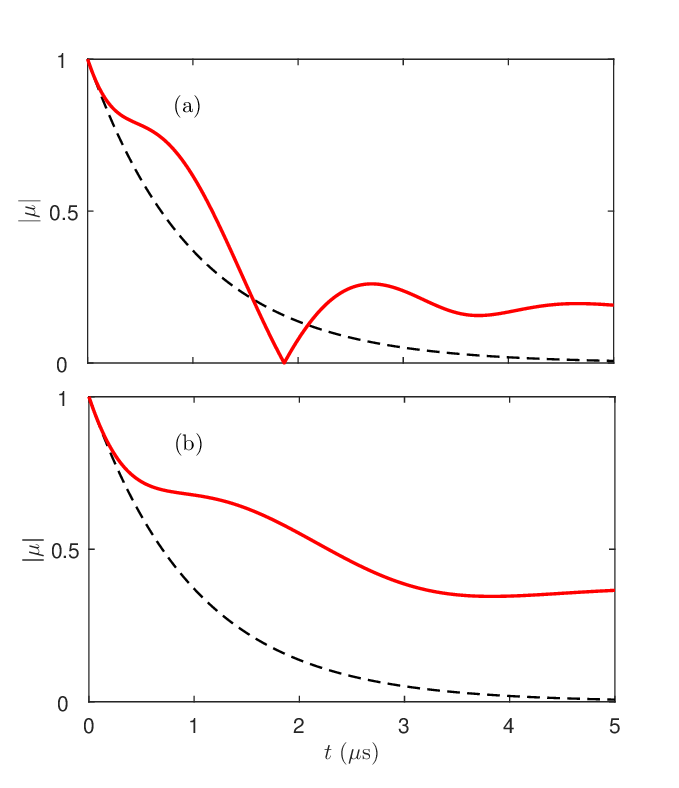}

\caption{Dynamic phases of the decohering evolution starting from the same
initial point $\alpha_{0}=1$ in the complex $(\alpha,\alpha^{\ast})$-plane
under positive detunings. The common parameters are equal to those
used in Fig.~\ref{fig:dynamic_phases}. The only differing parameters
are: (a) $\Delta_{d}/2\pi=520$~kHz and $\Omega/2\pi=100$~kHz;
(b) $\Delta_{d}/2\pi=220$~kHz and $\Omega/2\pi=100$~kHz.~\protect\label{fig:trajectory}}
\end{figure}

Finally, when the driving strength is strong (typically $\Omega>\Delta_{d}$),
the qubit accumulates energy faster than its dissipation into the
cTLS, resulting in a dynamic balance between driving and relaxation
that has a limit cycle radius even larger than the initial radius
$|\alpha_{0}|$. This near-resonant strong driving hence gives rise
to the amplifying phase IV.

\section*{APPENDIX E: Transient dynamics analysis}

Figure~\ref{fig:Q-dist_evolution} describes the evolution from the
initial distribution of a fully-inverted excited state, i.e. $\rho_{q}(0)=\left|e\right\rangle \left\langle e\right|$,
making $Q(\alpha,\alpha^{\ast},0)=\left|\alpha\right|^{2}\exp\left\{ -\left|\alpha\right|^{2}\right\} /\pi$.
Then, the evolution of the $Q$-distribution can be solved analytically
by Eq.~(\ref{eq:Q-Green}) as 
\begin{align}
Q(\alpha,\alpha^{\ast},t)= & \left[\frac{e^{-2\gamma\bar{\nu}t}\left|\alpha-\mu_{0}\right|^{2}}{4\sigma^{4}}+\frac{2\sigma^{2}-e^{-2\gamma\bar{\nu}t}}{2\sigma^{2}}\right]\nonumber \\
 & *\frac{1}{2\pi\sigma^{2}}\exp\left\{ -\frac{\left|\alpha-\mu_{0}\right|^{2}}{2\sigma^{2}}\right\} ,\tag{E1}\label{eq:Qini.E}
\end{align}
where $\mu_{0}=\mu_{\mathrm{ss}}\left[\exp(i\Delta_{d}t)-\exp(-\gamma\bar{\nu}t)\right]$.
The first term inside the square bracket corresponds to a ring-shaped
distribution which vanishes at long time limit $t\to\infty$. Whereas,
the second term corresponds to a disk-shaped distribution which vanishes
at $t=0$. The decay rate $2\gamma\bar{\nu}$ of the ring-shaped distribution
matches the growing rate of the disk distribution. We plot Fig.~\ref{fig:Q-dist_evolution}
to indicate these characteristics with the driving parameters $\Omega/2\pi=500~\textrm{kHz}$
and $\Delta_{d}/2\pi=-300~\textrm{kHz}$.

\begin{figure}
\includegraphics[bb=30bp 0bp 410bp 180bp,clip,width=8.5cm]{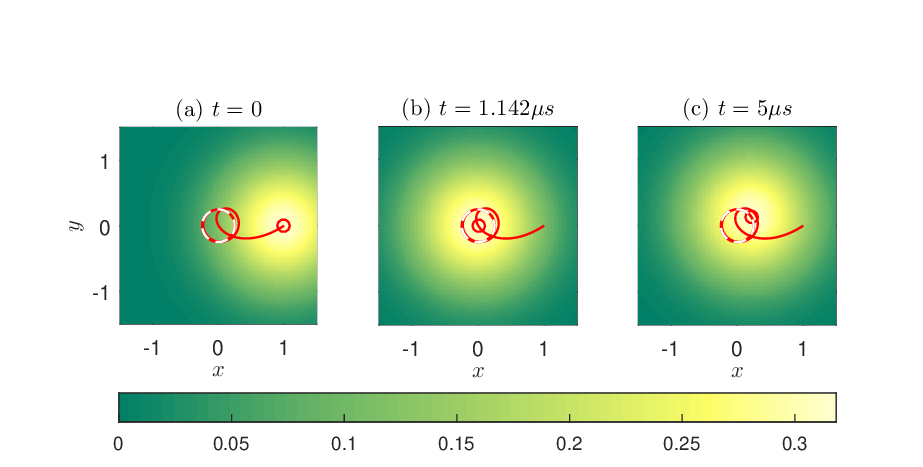}

\caption{The evolution of the qubit $Q$-distribution plotted as contours on
$XY$-quadrature plane, with the coherent state $\left|1\right\rangle $
as the initial state, shown as subplots at three typical time moments
(a) $t=0$, (b) $t=1.142\mu$s, and (c) $t=5\mu$s. The red curves
show the trajectory of the mean $\mu(t)$ and the small red circles
mark the locations of $\mu(t)$ at the these three moments. The overlay
white dashed curve indicates the limit cycle. The driving field parameters
are equal to those in Fig.~\ref{fig:dynamic_phases}(b).~\protect\label{fig:Q-evol-phase2}}
\end{figure}

If the initial state is rather a coherent state $\left|\alpha_{0}\right\rangle $,
the initial $Q$-distribution on the quadrature plane would be a Gaussian
disk about $\alpha_{0}$ away from the origin, as illustrated in Fig.~\ref{fig:Q-evol-phase2}(a)
with $\alpha_{0}=1$, instead of a ring shape about the origin. The
decoherence dynamics then mandates the trajectory of the mean $\mu(t)$
to converge to the limit cycle, while the distribution always remain
Gaussian. Given the operating phase to be type II, the mean reaches
the center first at $t=1.142\mu$s (shown in subplot (b)) before converging
towards the limit cycle of a radius smaller than $\alpha_{0}=1$ (shown
in subplot (c)). The variance of the Gaussian distribution is slightly
increased throughout the evolution.

\section*{APPENDIX F: Comparison of analytical and numberical solutions}

\begin{figure}
\includegraphics[bb=5bp 0bp 308bp 315bp,clip,width=8.5cm]{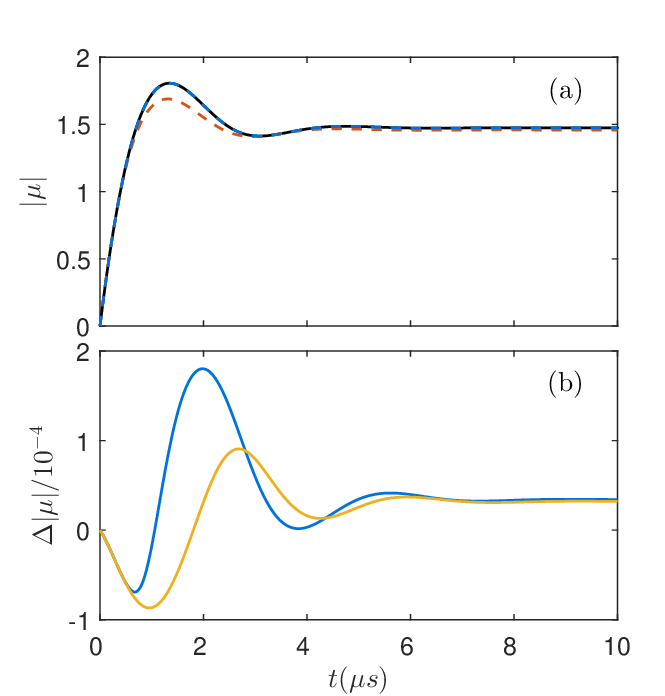}\caption{Evolutions of the distribution center $\mu(t)$ of the qubit density
matrix $\rho_{q}$, expressed by its $Q$-representation, solved using
the analytical as well as numerical approaches. The first excited
state $\left|e\right\rangle $ is assumed as the initial state and
physical parameters used are identical to those of Fig~\ref{fig:Q-dist_evolution}.
(a) shows the magnitude $|\mu(t)|$ where the black solid curve corresponds
to the analytical solution. The blue dashed curve (eclipsing the black
curve) corresponds to the numerical solution with 8 excited levels
are simulated while the orange dashed curve corresponds to that with
4 excited levels simulated. (b) shows the difference $\Delta|\mu(t)|$
of the numerical solutions from the analytical one. The blue curve
represents the scenario with 8 excited levels while the yellow represents
that with 19 excited levels.~\protect\label{fig:AnaVSNum}}
\end{figure}

To verify the accuracy of our analytical solution, we use matrix exponentiation
with Euler's equal step size to numerically solve Eq.~(\ref{eq:master_eqn})
in superoperator form and compare it with the analytical solution.
Figure~\ref{fig:AnaVSNum} shows the evolution of $|\mu(t)|$ with
full population at $\left|e\right\rangle $ as the initial state,
computed using both the analytical results of Sec.~III and the numerical
method, which adopts the cases where up to the lowest 4, 8, and 19
excited levels are considered as the computational Hilbert spaces.
Subplot (a) show the magnitude $|\mu|$ while subplot (b) shows the
difference $\Delta|\mu|$ between the analytical $|\mu|=|\mu_{ss}\exp(i\Delta_{d}t)+(\alpha_{0}-\mu_{ss})\exp(-\bar{\nu}\gamma t)|$
and the numerical $|\mu|$. The numerical $\mu(t)$ is identical to
$\left\langle a(t)\right\rangle =\textrm{tr}(a\rho_{q}(t))$ such
that given the numerical solution of $\rho_{q}(t)$ at time $t$ as
some $n\times n$ matrice, the operator $a$ is also expanded under
the same Fock basis of $n$ vectors, over which the trace is then
taken to compute a numerical value.

The identity between $\left\langle a(t)\right\rangle $ and $\mu(t)$
is recognized if we take the same trace but over the continuous coherent-state
space, i.e. $\left\langle a\right\rangle =\int d^{2}\alpha\left\langle \alpha\right|\rho_{q}a\left|\alpha\right\rangle /\pi=\int d^{2}\alpha\,\alpha Q(\alpha,\alpha^{\ast},t)$
where the $Q$ quasi-probability distribution is given by Eq.~(\ref{eq:Qini.E})
for $\left|e\right\rangle $ being the initial state. We recall that
$Q(\alpha,\alpha^{\ast},t)$ has two terms: one ring-shaped distribution
and one disk-shaped distribution. The former contributes, up to some
constant coefficent, a product of $\alpha|\alpha-\mu|^{2}$ and a
Gaussian function in the integrand; the latter contributes a product
of $\alpha$ and the same Gaussian. Writing $\alpha=x+iy$ and $\mu=A+iB$,
$\alpha|\alpha-\mu|^{2}$ contains four terms of the form $x(x-A)^{2}=(x-A)^{3}+A(x-A)^{2}$,
i.e. an odd power plus an even power about the axis $x=A$. Since
the Gaussian is itself even about $x=A$, when multiplied by this
cubic function, the $(x-A)^{3}$ term would vanish after integration
due to the symmetry in the integral while the $A(x-A)^{2}$ term would
contribute $A\sigma^{2}$ after integration by parts. The integration
over $y$ is about the Gaussian only and contributes only a unity
coefficient. Similarly, the term $iy(y-B)^{2}$ becomes $iB\sigma^{2}$
after integration. The rest two cross terms such as $iy(x-A)^{2}=i(y-B)(x-A)^{2}+iB(x-A)^{2}$
can be similarly integraded, contributing $iB\sigma^{2}$ and $A\sigma^{2}$
respectively. To summarize,
\begin{equation}
\frac{e^{-2\gamma\bar{\nu}t}}{4\sigma^{4}}\int_{-\infty}^{\infty}\alpha\left|\alpha-\mu\right|^{2}\mathcal{N}(\mu,\sigma^{2})d^{2}\alpha=\frac{e^{-2\gamma\bar{\nu}t}}{2\sigma^{2}}(A+iB),\tag{F1}
\end{equation}
where $\mathcal{N}(\mu,\sigma^{2})$ denotes the Gaussian. For the
product of $\alpha$ and the Gaussian, write $x=(x-A)+A$ and $iy=i(y-B)+iB$.
Thus following similar observations about evenness and oddities of
the integrands, we have
\begin{equation}
\frac{2\sigma^{2}-e^{-2\gamma\bar{\nu}t}}{2\sigma^{2}}\int_{-\infty}^{\infty}\alpha\mathcal{N}(\mu,\sigma^{2})d^{2}\alpha=\frac{2\sigma^{2}-e^{-2\gamma\bar{\nu}t}}{2\sigma^{2}}(A+iB).\tag{F2}
\end{equation}
Summing the two results give rise to $A+iB=\mu$, hence proving $\mu=\left\langle a\right\rangle $.

Using the same physical parameters as those of Fig~\ref{fig:Q-dist_evolution},
i.e. $\Delta_{d}/2\pi=-300$~kHz and $\Omega/2\pi=500$kHz, the plots
of Fig.~\ref{fig:AnaVSNum} demonstrate the accuracy of the analytical
approach, which is apparently more accurate than the numerical approach
when limited number of levels in $\rho_{q}$ is considered. Even when
the number of discrete levels is expanded to 20, the discrepany from
the numerical solution is on the order of $10^{-4}$, negligible compared
to the overall magnitude of $|\mu|$ over the entire evolution duration
considered.

\end{document}